\documentclass[a4paper,11pt]{article}
\usepackage{jheppub}
\usepackage{jheppub}
\usepackage{graphicx,amsfonts,amssymb,amsmath}
\usepackage{bm}
\title{Entanglement, Tensor Networks and Black Hole Horizons}

 \author{J. Molina-Vilaplana}
 \author{and J. Prior}
 \affiliation{Universidad Polit\'ecnica de Cartagena,\\C/Dr Fleming S/N 30202 Cartagena, Spain}

\emailAdd{javi.molina@upct.es}
\emailAdd{javier.prior@upct.es}

\abstract{We elaborate on a previous proposal by Hartman and Maldacena on a tensor network which accounts for the scaling of the entanglement entropy in a system at a finite temperature. In this construction, the ordinary entanglement renormalization flow given by the class of tensor networks known as the \emph{Multi Scale Entanglement Renormalization Ansatz} (MERA), is supplemented by an additional entanglement structure at the length scale fixed by the temperature. The network comprises two copies of a MERA circuit with a fixed number of layers and a pure matrix product state which joins both copies by entangling the infrared degrees of freedom of both MERA networks. The entanglement distribution within this \emph{bridge} state defines reduced density operators on both sides which cause analogous effects to the presence of a black hole horizon when computing the entanglement entropy at finite temperature in the AdS/CFT correspondence. The entanglement and correlations during the thermalization process of a system after a quantum quench are also analyzed. To this end, a full tensor network representation of the action of local unitary operations on the bridge state is proposed. This amounts to a tensor network which grows in size by adding succesive layers of bridge states.  Finally, we discuss on the holographic interpretation of the tensor network through a notion of distance within the network which emerges from its entanglement distribution.}


\begin{document} 
\maketitle
\flushbottom

\section{Introduction}
\label{sec:intro}
Since its initial formulation, the AdS/CFT correspondence \cite{adscftbib} has provided a huge amount of valuable knowledge related with the study of non-perturbative  effects in quantum field theory, despite a method to explicitly construct a bulk gravity theory from a boundary field theory without invoking its string theory roots, is still lacking \cite{douglas11}. Nowadays, it is also widely accepted a renormalization group interpretation of the correspondence, in which the renormalization scale becomes the extra radial dimension of the AdS duals while the beta functions of the boundary field theory are the saddle point equations of motion of the bulk gravity theory \cite{hologRG}. 

 Recently, the study of entanglement in strongly correlated quantum many body systems has provided a set of real space quantum renormalization group methods such as the density matrix renormalization group (DMRG) \cite{white92}, and the tensor network state (TNS) representations.  The later includes matrix 
products states (MPS) \cite{Cirac07}, projected entangled-pair states (PEPS) \cite{Verstraete09}, multi-scale entanglement renormalization ansatz (MERA) \cite{vidalmera}, tensor renormalization group (TRG) \cite{Levin07}, tensor-entanglement-filtering renormalization (TEFR) \cite{GuWen09} and algebraically contractible tensor network representations \cite{clark}. A tensor network description of the wavefunction of a quantum many body system is given by a collection of tensors whose values are determined by means of a variational search procedure. Those are connected into a network while the number of parameters defining them is much smaller than the dimension of the full system's Hilbert space. This allows an efficient representation of the wavefunction of the system in the thermodynamic limit. 

 In \cite{vidal}, tensor network states have been classified according to the connectivity of the sites within the network.  On one hand, there are networks that reproduce the physical connectivity between the sites as specified by the pattern of interactions dictated by the Hamiltonian. The MPS representation for one dimensional systems and its generalization for higher dimensional systems, lie within this category. Otherwise, there are tensor network representations in which the tensor connectivity organizes the quantum entanglement within that state along different length scales associated with successive coarse grained versions of the original lattice. These representations span an \emph{additional} dimension related with the RG scale which has led to define a generalized notion of holography inspired by the AdS/CFT correspondence \cite{swingle}. After the initial proposal, a substantial amount of work has appeared supporting and extending the original idea \cite{vidal, Malda_TNS, molina, matsueda, Taka_MERA, Molina11, vanram11, soliton, ads_mera_tak, cMERA, swingle12,sptmera13}.

 In this work, based on a previous proposal by Hartman and Maldacena \cite{Malda_TNS}, we elaborate on a tensor network which accounts for the scaling behaviour of  the entanglement entropy in a quantum system at finite temperature. There, authors depart from the thermofield double construction of the eternal black hole \cite{Israel, Malda_Eternal} to propose a tensor network which doubles the standard MERA for a pure state \cite{Malda_TNS, matsueda,Taka_MERA}. Here, we provide a detailed tensor network construction which implements such a proposal by introducing a bridge state between both copies of the MERA network. This network also accounts for the behaviour of the entanglement and correlations during the thermalization process of a system after a quantum quench. To this end, a tensor network representation for the action of local unitary operations on the bridge state is proposed. This amounts to a tensor network which grows in size by adding layers of these bridge states.  Under the action of these local operations, an initially highly localized entanglement distribution between sites located on both copies of the MERA network, scrambles in a way which may be efficiently represented by the network.  We also discuss on the holographic interpretation of the tensor network based on a notion of distance between its sites, which emerges from the entanglement structure it supports. Namely, it has been argued that quantum entanglement between separated regions of boundary system is a key to the emergence of a classical smooth spacetime geometry in the bulk \cite{VanRaam}. Thus, in the tensor network construction, each MERA copy correspond to the degrees of freedom lying in the two exterior regions of the eternal black hole. The future interior region of the eternal black hole, i.e, the entanglement between the degrees of freedom in both exterior regions according to \cite{VanRaam}, is then encoded in the pattern of entanglement supported by the bridge state.

This paper is organized as follows: in Section 2 we review the class of MPS and MERA tensor network states for gapped and critical one dimensional systems respectively, and discuss on the computation of entanglement entropy in these states, which has led to propose their holographic interpretation in \cite{swingle}. In Section 3, we elaborate on previous proposals \cite{Malda_TNS, matsueda} to provide a detailed tensor network description of a thermal state in AdS/CFT. The prescription is builded up in order to qualitatively account for the scaling of the entanglement entropy known for the AdS dual of a CFT state at a finite temperature. With this aim, we first introduce the tensor network representation of reduced density operators from which the bridge state is subsequently derived. In Section 4, we present a tensor network description of a local time evolution of the bridge state. According to \cite{Malda_TNS}, this addresses the behaviour of the entanglement and correlations during the thermalization process of a system after a quantum quench. Finally we discuss some issues concerning the holographic interpretation of the tensor network.

\section{Tensor Network states}
Tensor network states are a newly developed class of numerical real space renormalization group methods which efficiently describe ground states and low lying excitations of strongly correlated quantum systems. In this section we discuss on how a tensor network represents the entanglement structure of the wavefunction by reviewing two kinds of tensor network representations, MPS and MERA, and how entanglement entropy is computed within these states.

\subsection{Review of tensor network representations}
 The tensor network representation of the wavefunction of a one dimensional gapped system is known as Matrix Product State (MPS). We will briefly discuss on it by considering the MPS representation of a lattice with $N$ sites, each being described by vectors in a Hilbert space of dimension $d$. The Hilbert space of this system is normally spanned by the tensor product of local basis states $\vert j_1\rangle \otimes \vert j_2\rangle \otimes \cdots \otimes  \vert j_{N-1} \rangle \otimes \vert j_N\rangle= \vert j_1 j_2 \cdots j_N\rangle$,  which amounts to a description of the wavefunction in a total Hilbert space of dimension $d^N$ which is given by, 

\begin{equation}
\label{eqp0}
\vert \Psi\rangle=\sum_{j_1,j_2, \cdots ,j_N=1}^d \mathcal{T}_{j_1,j_2, \cdots ,j_N} \vert j_1 j_2 \cdots j_N\rangle.
\end{equation}

The MPS \emph{ansatz} \cite{Cirac07} assumes that each probability amplitude $\mathcal{T}_{j_1,j_2, \cdots ,j_N}$ can be written in terms of a set of matrices as,%
\begin{equation}
\label{eqp1}
\vert\Psi \rangle=\sum_ {j_1,j_2,...,j_N=1}^d \mathcal{F} \left( A^{[1]}_{j_1} A^{[2]}_{j_2} \cdots A^{[k]}_{j_k} A^{[k+1]}_{j_{k+1}} \cdots A^{[N]}_{j_N}\right) \, \vert j_1 j_2 \dots j_N\rangle,
\end{equation}
 where $A^{[k]}_{j_k}$ is a set of $d$ complex matrices of dimension $(\chi_{k-1}\times \chi_k)$ labeled by the physical index $j_k$ and $\mathcal{F}(.)$ is a function which maps the matrix $A^{[1]}_{j_1} A^{[2]}_{j_2} \cdots A^{[N]}_{j_N}$ of dimension $(\chi_0 \times \chi_N)$ into the scalar $\mathcal{T}_{j_1,j_2, \cdots ,j_N}$. If one deals with periodic boundary conditions, $\mathcal{F}(.)$ amounts to the trace of the resulting matrix product, i.e, ${\rm Tr}\left( A^{[1]}_{j_1} A^{[2]}_{j_2} \cdots A^{[N]}_{j_N}\right) $. In case of having open boundary conditions, the map is implemented by introducing a left vector $\langle \Phi_0\vert$ and a right vector $\vert \Phi_N\rangle$ to obtain $\mathcal{T}_{j_1,j_2, \cdots ,j_N} = \langle \Phi_0 \vert A^{[1]}_{j_1} \cdots A^{[N]}_{j_N} \vert \Phi_N \rangle$.
 
 The MPS representation has a gauge freedom which may be fixed at any site of the lattice. This fixing is related with the amount of block-wise entanglement in the state. Let us clarify this point by sketching how to exploit this freedom to transform an unconstrained matrix product representation into a canonical form made up of a series of contiguous Schmidt decompositions. To proceed, let us first to insert a non-singular square matrix $\Lambda_k $ and its inverse $\Lambda_k^{-1}$, both with dimensions $(\chi_k \times \chi_k)$ at site $k$. We impose $\Lambda_k$ to be diagonal with real elements $\lambda_{\alpha_k}$. It is straightforward to note that, by changing $ A^{[k]}_{j_k} A^{[k+1]}_{j_{k+1}} \to A^{[k]}_{j_k}\Lambda_k\, \Lambda_k^{-1} A^{[k+1]}_{j_{k+1}}$ into the full matrix product, one leaves invariant the MPS representation of the amplitude $\mathcal{T}_{j_1,j_2, \cdots ,j_N}$, which now reads as,
\begin{equation}
\label{eqp2}
\langle \Phi_0 \vert A^{[1]}_{j_1} \cdots A^{[k]}_{j_k}\, \Lambda_k\,  \Gamma^{[k+1]}_{j_{k+1}} \cdots A^{[N]}_{j_N} \vert \Phi_N \rangle,
\end{equation}
 where $\Gamma^{[k+1]}_{j_{k+1}} = \Lambda_{k}^{-1}\, A^{[k+1]}_{j_{k+1}}$. Using the identity $\Lambda_k=\sum_{\alpha_k=1}^{\chi_k}\, \lambda_{\alpha_k}\, \vert \alpha_k\rangle \,  \langle \alpha_k \vert $, it is possible to write the state $\vert \Psi \rangle$ as,
\begin{equation}
\label{eqp3}
\vert \Psi \rangle = \sum_{j_{[1\cdots N]}}\, \sum_{\alpha_k=1}^{\chi_k}\, \lambda_{\alpha_k} \langle \Phi_0 \vert A^{[1]}_{j_1} \cdots A^{[k]}_{j_k}\vert \alpha_k\rangle \, \langle \alpha_k \vert \Gamma^{[k+1]}_{j_{k+1}} \cdots A^{[N]}_{j_N} \vert \Phi_N \rangle \, \vert j_{[1 \cdots k]}\rangle \vert j_{[k+1\cdots N]}\rangle,  
\end{equation}
or in a more compact form as,
 \begin{equation}
\label{eqp4}
 \vert \Psi \rangle = \sum_{\alpha_k=1}^{\chi_k}\, \lambda_{\alpha_k}\, \vert \phi_{\alpha_k}^{[L]}\rangle\, \vert \phi_{\alpha_k}^{[R]}\rangle,
\end{equation}
where,
 \begin{eqnarray}
\label{eqp5}
 \vert \phi_{\alpha_k}^{[L]}\rangle &=& \sum_{j_{[1\cdots k]}}\, \langle \Phi_0 \vert A^{[1]}_{j_1} \cdots A^{[k]}_{j_k}\vert \alpha_k \rangle \,  \vert j_{[1 \cdots k]}\rangle \\ \nonumber
 \vert \phi_{\alpha_k}^{[R]}\rangle &=&  \sum_{j_{[k+1\cdots N]}}\, \langle \alpha_k \vert \Gamma^{[k+1]}_{j_{k+1}} \cdots A^{[N]}_{j_N}\vert \Phi_N\rangle \,  \vert j_{[k+1 \cdots N]}\rangle, 
\end{eqnarray}
are states of the blocks comprising all the sites to the left-$[L]$ and all the sites to the right-$[R]$  of site $k$ respectively. In order to identify the bipartite splitting of the state in Eq.(\ref{eqp4}) as a genuine Schmidt decomposition of $\vert \Psi \rangle$, it is necessary to impose that $\lbrace \vert \phi_{\alpha_k}^{[L]}\rangle \rbrace$ and $\lbrace \vert \phi_{\alpha_k}^{[R]}\rangle \rbrace$ both constitute $\chi_k$-dimensional basis for the states of the $[L]$-block and the $[R]$-block respectively. This may be accomplished by imposing a set of orthogonality constraints on all the matrices $A^{[m]}_{j_m}$, $1 \leq m \leq N$, where  $A^{[k+1]}_{j_{k+1}} \equiv \Gamma^{[k+1]}_{j_{k+1}}$. A detailed discussion of these constraints may be found in, for instance, \cite{Cirac07}. Finally, one would require $\vert \Psi \rangle$ to be normalised, which amounts to ${\rm Tr}(\Lambda_k^2) = 1 = \sum_{\alpha_k=1}^{\chi_k}\, \lambda_{\alpha_k}^{2} = 1$, allowing the diagonal matrix $\Lambda_k$ to be identified with the matrix of the Schmidt coefficients and $\chi_k$ with the Schmidt rank of the decomposition Eq.(\ref{eqp4}).

It is thus straightforward to see how the dimension of $\Lambda_k$ measures the amount of entanglement between these blocks. To this aim, we compute the entanglement entropy between the two blocks,
 \begin{equation}
\label{eqp6}
 S(\rho_{L, (R)})=-{\rm Tr}\, (\rho_{L, (R)}\,\log \rho_{L, (R)})= -\sum_{\alpha_k=1}^{\chi_k}\, \lambda_{\alpha_k}^{2}\, \log \lambda_{\alpha_k}^{2},
\end{equation}
where $\rho_{L, (R)}={\rm Tr}_{R,(L)} \vert \Psi \rangle \langle \Psi \vert$ are given by,
\begin{eqnarray}
\label{eqp7}
\rho_L &=&\sum_{\alpha_k =1}^{\chi_k} \lambda_{\alpha_k}^2 \vert\phi_{\alpha_k}^{[L]}\rangle \langle\phi_{\alpha_k}^{[L]}\vert \\ \nonumber
\rho_R &=&\sum_{\alpha_k=1}^{\chi_k} \lambda_{\alpha_k}^2  \vert\phi_{\alpha_k}^{[R]}\rangle \langle\phi_{\alpha_k}^{[R]}\vert.
\end{eqnarray}
 
In the tensor network literature, $\Lambda_k$ represents an entangled bond between the $[L]$-states and the $[R]$-states. From Eq. (\ref{eqp6}) it is easy to provide an upper bound to the entanglement between the adjacent blocks $[L]$ and $[R]$, which is maximal when $\lambda_{\alpha_k}=1/\sqrt{\chi_k}$, $\forall\, \alpha_k$, in which case $S(\rho_{L,(R)})= \log \chi_k$. 
  
 The above procedure yields an orthonormalised matrix product representation of $\vert \Psi \rangle$ equivalent to a particular Schmidt decomposition which depends on where the left and right bipartion occurs. It is then possible to repeat the procedure along each site in the lattice to obtain (by impossing another set of constraints \cite{Cirac07}) a canonical MPS representation given by,
 \begin{equation}
\label{eqp8}
\mathcal{T}_{j_1 \cdots j_N}=\langle \Phi_0 \vert \Gamma^{[1]}_{j_1}\Lambda_1 \cdots \Gamma^{[k]}_{j_k}\, \Lambda_k\,  \Gamma^{[k+1]}_{j_{k+1}} \cdots \Lambda_{N-1}\Gamma^{[N]}_{j_N} \vert \Phi_N \rangle,
\end{equation} 
which explicitly shows how a tensor network given by a set of $\Gamma$ and $\Lambda$ matrices, encodes the block-wise entanglement structure of the wavefunction\footnote{An alternative way to find the canonical form of an MPS representation can be found in \cite{tebd}}.

\begin{figure}[t]
\label{figure1}
\centerline{\includegraphics[width=2.5 in]{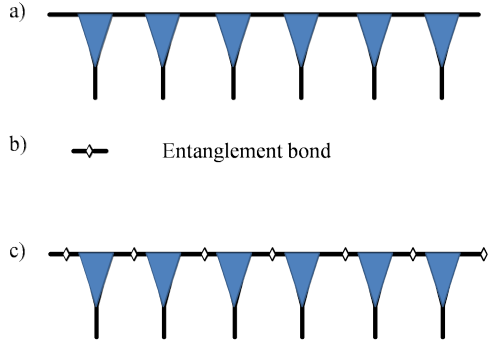}}
 \centering \caption{Representation of an MPS state. a) Pictorial representation of Eq.(\ref{eqp1}) in which each blue triangle corresponds to a matrix $A^{[m]}_{j_m}$. Downward legs represent physical indices $j_m $ while horizontal lines refer to an index contraction between adjacent matrices. b) An entangled bond between two nearby sites given by a matrix $\Lambda$ is depicted as a diamond within an horizontal line. c) Pictorial representation of Eq.(\ref{eqp8}) in which blue triangles correspond to matrices $\Gamma$. Downward legs represent physical indices $j$ and horizontal lines with an inserted diamond are entangled bonds $\Lambda$ between nearby sites.}
\end{figure} 

\subsection{Entanglement renormalization tensor networks}
The entanglement renormalization (MERA) tensor network representation of a one-dimensional state takes up a decomposition of $\mathcal{T}_{i_1,i_2,\cdots, i_N}$ in terms of a set of tensors organized in a two-dimensional layered graph. Each site of the graph represent a tensor and those may divided into the type $w\,- \mbox{\tiny{$\left(\begin{array}{c} 2 \\2  \end{array}\right)$}}$ unitary tensors known as \emph{disentanglers} and  the type $\lambda\, - \mbox{\tiny{$\left(\begin{array}{c} 1 \\2  \end{array}\right)$}}$  tensors called \emph{isometries} (Figure 2). In a quantum critical system, a MERA tensor network posses a characteristic scale invariant structure in which a unique $w$ and a unique $\lambda$ define the full MERA graph.  

The MERA representation of the wavefunction implements an efficient real space renormalization group procedure through a tensor network organized in different layers labelled by $u$, where $u=0$ for the state lying on the initial lattice (UV). Each layer of MERA performs a renormalization transformation in which, prior to the coarse graining of a block of typically two sites located at layer $u$ into a single site by means of a $\lambda^{u}$ tensor, the short range entanglement between these sites is removed through the  action of a {\em disentangler} $w^{u}$. Thus, each layer of the MERA network, decouple the relevant low energy degrees of freedom from the high energy ones, which are then removed, by unitarily transforming with \emph{disentanglers} small regions of space. As one iteratively proceeds, the coarse-graining transformation carried out by MERA, generates an RG map that can be applied arbitrarily many times ($u \to \infty$ for the case of an infinite scale invariant system). 

In \cite{swingle}, it was firstly observed that, from the entanglement structure of an static wavefunction represented by MERA, one may define a higher dimensional geometry in which, apart from the coordinates $\vec{x}$ labelling the sites in the lattice, it is possible to add the "radial" coordinate $u$ which accounts for the hierarchy of scales. The discrete geometry emerging at the critical point is a discrete version of the hyperbolic AdS spacetime,
\begin{eqnarray}
ds^2 = R^{2}\left(du^2 + e^{-2\, u}d\vec{x}^2\right)~,
\label{AdS}
\end{eqnarray}
  where $R$ is a constant called the AdS radius; it has the dimension of a length and it is related with the curvature of the AdS space. With this choice of the spacetime coordinates, the one dimensional quantum critical system lies at the boundary ($u = 0$) of the MERA geometry. 
  
\begin{figure}[t]
\label{fig1.eps}
\includegraphics[width=3.25 in]{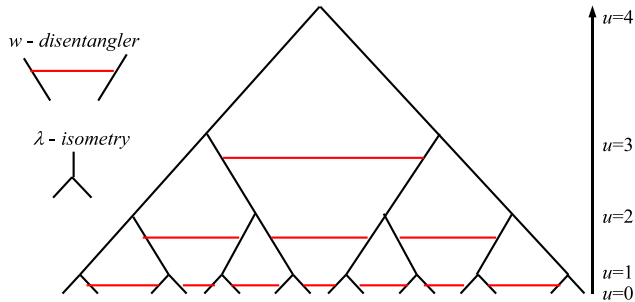}
 \centering \caption{Scale invariant MERA network representing a CFT.}
\end{figure}

\subsection{Entanglement Entropy in tensor networks states}
Following \cite{vidal}, an upper bound for the entanglement entropy $S_A$ in a tensor network state can be given as follows. First, one arbitrarily splits the network into two connected parts, $\Omega_A$ and $\Omega_B$, where $A \subset \Omega_A$ and $\Omega_B$ contains the rest of sites of the network. The region $\Omega_A$ amounts to the tensor network description of the reduced density matrix $\rho_A$ of the subsystem $A$. Then, we count the $n(A)$ bond indices which connect the regions $\Omega_A$ and $\Omega_B$. Since each bond contributes a maximum of $\log\, \chi$ to the entropy of $\rho_A$, an upper bound to $S_A$ can be written as,
\begin{equation}
S_A \leq n(A) \log \, \chi.
\label{eq1.1}
\end{equation}

As a result, the splitting which minimizes $n(A)$ provides the tightest upper bound to $S_A$ and the entanglement entropy $S_A$ scales proportional to $n(A)$.  In the discrete "geometry" of the tensor network, $n(A)$ acts as a "distance" measuring the size of the boundary $|\partial \Omega_A|$ of the the region $\Omega_A$. Thus is possible to recast Eq. (\ref{eq1.1}) as, 
\begin{equation}
S_A \approx |\partial\Omega_A|\log \, \chi,
\label{eq1.2}
\end{equation}
which amounts to say that the entropy $S_A$ is proportional to the length of the boundary of region $\Omega_A$. The scaling of $S_A$ for an specific tensor network state, is obtained as one states the explicit dependence of $|\partial\Omega_A|$  with the size $L$ of region $A$. In a 1-dimensional MPS, $|\partial \Omega_A|=2$ while in the discretized hyperbolic geometry of a 1-dimensional scale invariant MERA, $|\partial \Omega_A| \sim \log\, L$.

The standard way to display the AdS/MERA connection \cite{swingle} compares the computation of the entanglement entropy in both cases. In the classical gravity limit of AdS/CFT, Ryu and Takayanagi (RT) derived a formula to obtain the entanglement entropy of a region $A$ provided that the (boundary) conformal
field theory admits an holographic gravity dual \cite{ryu}. In the RT proposal, the entanglement entropy is obtained through the computation of a minimal surface in the dual higher dimensional gravitational geometry. As a result, $S_A$ is given by,
\begin{eqnarray}
S_{A}=\frac{{\rm Area}(\gamma_{A})}{4G^{(D+1)}_N}~,
\label{arealaw}
\end{eqnarray}
where $D$ is the number of spacetime dimensions of the boundary CFT, $\gamma_{A}$ is the $(D-1)$-dimensional static minimal surface in AdS$_{D+1}$ and $G^{(D+1)}_{N}$ is the $D+1$ dimensional Newton constant\footnote{In this paper, we will be mainly focused in the  $D=2$ case, for which the equation (\ref{arealaw}) reduces to, $S_{A}={\rm Length}(\gamma_{A})/4G^{(3)}_N$}. To look for the minimal surface $\gamma_{A}$ which optimally separates the degrees of freedom within $A$ from those lying in the complementary region $B$, amounts to obtain the severest entropy bound on the information hidden in the AdS$_{D+1}$ region related with $B$.

One may notice the close similarity between the formula (\ref{arealaw}) and the Eq.(\ref{eq1.2}) which gives the entanglement entropy of a region $A$ in MERA, as one realizes that $n(A) \sim |\partial\, \Omega_A|$ can be regarded as the minimal curve $\gamma_A$ in the RT proposal \cite{swingle}.

\section{Entanglement renormalization tensor networks and black holes}
\label{sec:black hole}
In this section we propose a simple MERA tensor network to describe quantum critical systems at finite temperature. In the AdS/CFT, the gravity dual of a finite temperature state is the well-known AdS black hole \cite{adscftbib}. The inverse temperature of the system is related with the distance of the boundary CFT to the black hole horizon. It has been argued that, inspired by the thermofield double construction of the eternal black hole \cite{Israel,Malda_Eternal}, a MERA network with an horizon may be described by doubling the standard MERA for a pure state and then connecting together the infrared regions of both networks \cite{Malda_TNS, matsueda,Taka_MERA} through a gluing-through-entanglement operation \cite{VanRaam}. In \cite{Taka_MERA} it is noted that the definition of this IR entangled state is rather ambiguous. Here, we show that the gluing operation may be satisfactorily characterized through the tensor network representations of density operators (MPDO) \cite{mpdo} and their \emph{purifications}. 

\subsection{Matrix product density operators and purifications}
\label{susec:mpdo}
 The standard way of building up an MPDO takes advantadge of  the fact that every mixed state can be seen
as the state of a partial subsystem of a bigger pure system, namely its \emph{purification}. If one models this purification as an MPS $\vert \Psi\rangle$, then the MPDO $\rho$ is obtained by tracing over the purifying degrees of freedom. In this picture the purification is builded by attaching to every original degree of freedom $i$, a locally accompanying purifying site $j$, in such a way that in the pure state, these pairs of sites correspond to one bigger site $ij$:
\begin{eqnarray}
\label{eq2.1}
\vert \Psi\rangle & = & \sum_{i_1,i_2,\cdots j_1,j_2,\cdots = 1}^{d}\, {\rm Tr}\left(A^{[1]}_{i_1 j_1}\, A^{[2]}_{i_2 j_2}\, \cdots \right)\vert i_1\rangle \vert j_1 \rangle \otimes \vert i_2\rangle \vert j_2 \rangle \otimes \cdots , \\ \nonumber
\rho & = & {\rm Tr}_{j_1,j_2,\cdots} \left(\vert \Psi \rangle \langle \Psi \vert \right)= \\ \nonumber 
&=& \sum_{i_1,i_2,\cdots i^{'}_1,i^{'}_2,\cdots = 1}^{d}\,{\rm Tr}\left( \underbrace{\sum_{j_1 j^{'}_1}A^{[1]}_{i_1 j_1}\otimes \bar{A}^{[1]}_{i^{'}_1 j^{'}_1}}_{M^{[1]}_{i_1 i^{'}_1}}\, \underbrace{\sum_{j_2 j^{'}_2}A^{[2]}_{i_2 j_2}\otimes \bar{A}^{[2]}_{i^{'}_2 j^{'}_2}}_{M^{[2]}_{i_2 i^{'}_2}}\, \cdots \right) \vert i_1\rangle \langle i^{'}_1 \vert \otimes \vert i_2\rangle \langle i^{'}_2 \rangle \otimes \cdots , \\ \nonumber 
&=&  \sum_{i_1,i_2,\cdots i^{'}_1,i^{'}_2,\cdots = 1}^{d}\,{\rm Tr}\left(M^{[1]}_{i_1 i^{'}_1}\, M^{[2]}_{i_2 i^{'}_2} \cdots \right) \vert i_1\rangle \langle i^{'}_1 \vert \otimes \vert i_2\rangle \langle i^{'}_2 \rangle \otimes \cdots ,
\end{eqnarray}
 where $\bar{A}$ denotes complex conjugation. One case which will be of interest to us in the following is the purification of the infinite temperature mixed state. Let us consider a block of $N$ $i$-sites in the completely mixed state,
 \begin{equation}
 \label{eq2.2}
 \rho_{N}=\frac{1}{d^{N}}\, \left( \mathbb{I}_{d}\right) ^{\otimes N}.
 \end{equation} 
 
 The entropy of $\rho_{N}$ is given by,
\begin{eqnarray}
\label{eq2.5}
S(\rho_{N})&=& S\left( \underbrace{\frac{1}{d}\,  \mathbb{I}_{d}\otimes \frac{1}{d}\,  \mathbb{I}_{d} \cdots \otimes \frac{1}{d}\,  \mathbb{I}_{d}}_{N}\right) = \underbrace{S\left( \frac{1}{d}\,  \mathbb{I}_{d}\right) +\cdots + S\left( \frac{1}{d}\,  \mathbb{I}_{d}\right) }_{N} = N\, \log d.
\end{eqnarray}
 
As $\rho_{N}$ factorizes, one may purify the local mixed state on each $i$-site in terms of a maximally entangled state at each $ij$-site. Indeed, noticing that,
\begin{equation} 
\label{eq2.3}
\frac{1}{d}\, \mathbb{I}_{d} = \sum_{\alpha=1}^{d}\frac{1}{d}\vert \alpha \rangle_{i} \langle \alpha \vert = {\rm Tr}_{j}\, \left[ \left( \sum_{\alpha=1}^{d}\frac{1}{\sqrt{d}}\vert \alpha \rangle_{i} \vert \alpha \rangle_{j} \right) \left( \sum_{\alpha'=1}^{d}\frac{1}{\sqrt{d}}\langle \alpha' \vert_{i} \langle \alpha' \vert_{j} \right)  \right], 
\end{equation}
it is easy to see that the purification is given by a maximally entangled state on each $ij$-site (which contributes $\log d$ to the entanglement entropy),
\begin{equation}
\label{eq2.4}
\vert \psi \rangle_{ij} = \sum_{\alpha=1}^{d}\frac{1}{\sqrt{d}}\vert \alpha \rangle_{i} \vert \alpha \rangle_{j}. 
\end{equation}

This amounts to write the global purified state as $\vert \Psi\rangle=(\vert \psi \rangle_{ij})^{\otimes N}$ which is an MPS of dimension 1\footnote{In the following we use the notation $\lbrace i \rbrace = \lbrace i_1 \cdots i_N\rbrace$, $\lbrace j \rbrace = \lbrace j_1 \cdots j_N\rbrace$ and $\lbrace i j\rbrace = \lbrace i \rbrace \cup \lbrace j \rbrace$. In case $d=2$, the state can be written as $\vert \Psi \rangle = \left( \frac{1}{\sqrt{2}}\, (\vert 0 0 \rangle  + \vert 1 1 \rangle)\right)^{\otimes N} $, i.e the purified MPS state resembles a system of $N$ Bell pairs shared between the $\left\lbrace  i \right\rbrace $ and $\left\lbrace  j \right\rbrace$ sites.}. In terms of matrices  one also may express  $\vert \Psi \rangle$ through $A_{ij}=\delta_{ij}$. 

\begin{figure}[t]
\label{MERA_FT1}
\includegraphics[width=4.75 in]{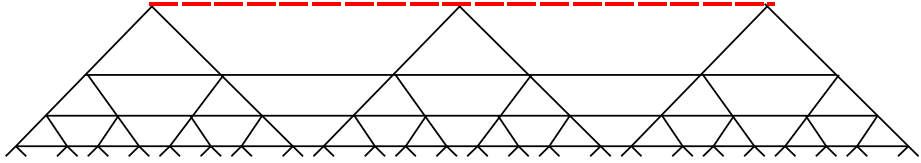}
 \centering \caption{Capped MERA network with $\log \, \beta/2\pi$ layers of tensors associated 
with the neighbouring critical point and an MPDO acting as a capping top layer (dashed red line). From now on we use the compressed pictorial notation in \cite{Malda_TNS} for the MERA network in which each vertex represents a tensor with five indices.}
\end{figure}

\subsection{A first ansatz}
\label{susec:firstansatz}
Following \cite{swingle} and \cite{matsueda, molina} we assume some general features for a tensor network \emph{avatar} of an AdS black hole  with an inverse temperature $\beta$: the system under consideration is gapped so first, we initially expect a region of discrete AdS geometry (standard MERA layers with tensors associated to the neighbouring critical point) for energy scales much greater than the temperature. As more MERA renormalization steps are carried out (as $u$ grows), it is expected that thermal effects gradually prevail on the IR sector of the system so, the coarse-grained description of the sites must incorporate these thermalized degrees of freedom. Eventually, after $\sim \log \, \beta$ renormalization steps, a final scale is reached. At this final scale, the reduced density matrix of any coarse grained site is proportional to the identity, and the coarse-grained density matrix completely factorizes. 

 In \cite{swingle}, this situation was interpreted as corresponding to the presence of a black hole horizon because: i) the geometry ends from the point of view of an observer "hovering" at fixed scale \footnote{In the gravitational terminology this is a fiducial observer FIDO for which the local temperature diverges near the horizon}, ii) the completely mixed state amounts to considering a thermal state with an infinite temperature, and the local temperature probed by a hovering observer diverges at simple black hole horizons, iii) the final layer has nonzero entropy because the coarse-grained sites are in mixed states. In particular, one expects that the entropy of a large block in the UV (first MERA layer) consists of two pieces: the contribution coming from the MERA \emph{curtain} plus an extensive piece due to the "horizon". 

 According to these arguments, one might naively guess the following ansatz for the MERA network of a finite temperature state (Figure 3): we build an hybrid tensor network composed by a finite number $u=\log \, \beta/2\pi$ of MERA layers and a matrix product density operator (MPDO) acting as a cap "horizon" layer. Each site in the top MPDO represents a cluster of $\sim e^{u}$  coarse grained sites of the original lattice. 

 As in \cite{swingle, molina}, we compute the entanglement entropy $S_A$ of a region $A$ with $|A|=\ell$ sites in the tensor network, to check if it matches some qualitative aspects of the entropy obtained through the AdS dual of a finite temperature state. To this end, we split the $S_A$ into two contributions: the first one, $S_A^{UV}$, corresponds to a MERA \emph{curtain} with $\log\, \beta/2\pi$ layers and a second one, $S_A^{IR}$, has a value which is controlled by the entropy of $\sim \ell/\beta$ sites in the top MPDO state. Namely, the top layer is a completely mixed state such as the one discussed in Eq. (\ref{eq2.2}). Thereby, the $S_A^{UV}$ amounts to count the number of bonds connecting $\Omega_A$ with the rest of the sites to give,
 \begin{equation}
 \label{eq2.5}
  S_A^{UV}\sim n(A)^{UV}\,\log\, \chi  =  \left( \underbrace{2+ \cdots + 2}_{\log\, \beta/2\pi}\right) \log\, \chi = 2\log\, \chi\, \log\, \frac{\beta}{2\pi},
 \end{equation}
 with $\chi$ the bond dimension of the MERA network. Eventually, once the top layer is reached, the original $\ell$ degrees of freedom  have been coarse grained into $\sim \ell\, e^{-u} = 2\pi\, \ell/\beta$ sites with dimension $d \sim \chi$. Thus, the extensive contribution $S_A^{IR}$ of these sites can be written as,
 \begin{equation}
 \label{eq2.6}
  S_A^{IR} = 2\pi\, \log\, \chi \, \left( \frac{\ell}{\beta}\right) ,
 \end{equation} 
 yielding a total amount of entanglement,
  \begin{equation}
 \label{eq2.7}
   S_A = S_A^{UV} + S_A^{IR} \sim  2\log\, \chi \, \left[ \frac{\pi \ell}{\beta} +  \log\,  \frac{\beta}{2\pi}\right].
 \end{equation}

 These general features of the tensor network for a finite temperature state, match the behaviour of the holographic entanglement entropy corresponding to the AdS$_3$ black hole background. The minimal length curves connecting two points $u,\, v$ located at the boundary of the AdS$_3$, separated by a distance $\ell=\vert u - v\vert$ are given by,
\begin{equation}
\label{eq2.8}
{\rm Length}(u,v) =  2 R\, \log \left[\frac{\beta}{\pi \epsilon} \sinh\left(\frac{\pi \ell}{\beta} \right)  \right]~,
\end{equation}
with $\epsilon$ the regularizing UV cut-off and $R$ the radius of AdS$_3$. Using the RT formula and $c=3\, R/2G_N^{(3)}$ \cite{brownhen}, one gets the entanglement entropy of a large block $A$ of length $\ell=\vert u-v\vert \gg \beta$,
\begin{eqnarray}
\label{eq2.9}
S_{A}=\frac{{\rm Length}(u,v)}{4G_N^{(3)}}=\frac{c}{3} \log \left[\frac{\beta}{\pi \epsilon} \sinh\left(\frac{\pi \ell}{\beta} \right)  \right]
\approx  \frac{c}{3}\, \frac{\pi \ell}{\beta} + \frac{c}{3} \log  \frac{\beta}{2 \pi \epsilon}.
\end{eqnarray}

 As expected, when the size of the interval $A$ is bigger than the distance of the horizon from the boundary, the geodesics probe the black hole horizon extending tangentially to it; this originates the extensive contribution to the entanglement entropy which describes  a thermal state at temperature $T=2\pi /\beta$. As a result, one realizes that Eq. (\ref{eq2.7}) and Eq.(\ref{eq2.9}) qualitatively agree once provided the fixing of the bond dimension given by,
 \begin{equation}
 \label{eq2.10}
 \log\, \chi = \frac{c}{6}.
 \end{equation}

\subsection{Doubled MERA networks and bridge states}
  Despite the picture given above yields anappealing result, the construction is rather unnatural from a tensor network point of view; since each layer in MERA  represents a different renormalized coarse grained version of the UV pure state, it is difficult to justify a capping layer given by a mixed state, in our case, the completely mixed state. A more crucial point is that the entropy gathering in the IR is not related with any bond counting process such as the one considered in, for instance, \cite{molina} to deal with a different kind of gapped systems.
 
  To overcome this difficulty and inspired by the thermofield double representation of the eternal black hole \cite{Malda_Eternal, matsueda, Malda_TNS}, a doubled MERA circuit representing the wavefunction of the thermofield double is proposed. The tensor network represents a gapped state with an scale invariant UV region. The circuit is composed of two copies of a MERA \emph{curtain} with $u =\log \, \beta/2 \pi$ layers, so the scale invariance is broken by the temperature $T=2\pi/\beta$. At this scale, we place a "bridge" MPS state which entangles the IR degrees of freedom of the two MERA \emph{curtains} (Figure 4).  
  
  This pure IR state is builded by attaching to every $i$-coarse grained IR degree of freedom corresponding to one copy, a locally purifying  $ j $-site corresponding to a coarse grained IR site of the other copy. One might say that the bridge MPS, in some sense, glues the two MERA curtains. The entanglement structure and distribution of the bridge MPS thus defines the infrarred MPDO  states corresponding to each MERA copy as discussed in section (\ref{susec:mpdo}). This construction suggests that the ordinary entanglement renormalization flow given by MERA has to be supplemented by additional entanglement structures at scales larger than $\sim \log\beta$.
  
  We choose to construct the bridge state using a set of simple building blocks which amounts to a collection of $N$ maximally entangled Bell pairs,
  \begin{equation}
\label{eq2.11}
\vert \psi \rangle_{ij} = \sum_{\alpha=1}^{\chi}\frac{1}{\sqrt{\chi}}\vert \alpha \rangle_{i} \vert \alpha \rangle_{j},
\end{equation}
 which yields the state,
 \begin{equation}
\label{eq2.11b}
\vert \Psi\rangle=(\vert \psi \rangle_{ij})^{\otimes N}.
\end{equation}

 It is worth to note that for every MPS state one can always construct a so-called \emph{parent} Hamiltonian. This is a local, frustration free, Hamiltonian which has the MPS state as its unique ground state and is gapped \cite{Cirac07}. Parent Hamiltonians are extremely useful as they allow us to use the MPS formalism to analyze the behaviour of one-dimensional Hamiltonians. The parent Hamiltonian of the state given in Eq. (\ref{eq2.11b}) may be generically written as,
 \begin{equation}
\label{eq2.11c}
H_{parent}=H_{\left\lbrace i  \right\rbrace} + H_{\left\lbrace j \right\rbrace} + H^{int}_{\left\lbrace i \, j \right\rbrace},
\end{equation} 
 where $H_{\left\lbrace i  \right\rbrace}$ and $H_{\left\lbrace j  \right\rbrace}$ are Hamiltonians acting only on  $\left\lbrace i  \right\rbrace$-sites and $\left\lbrace j  \right\rbrace$-sites respectively and $H_{\left\lbrace i\, j  \right\rbrace}^{int}$ mediates the interaction between the IR degrees of freedom of the two separated MERA \emph{curtains} from which the entanglement structure supported by the bridge state arises.

 Within this tensor network structure, the computation of the entanglement entropy $S_A$ of a block $A$ of $\ell$ sites in the first layer of one of the MERA curtains comes as follows: as before, we count the number $n(A)$ of bonds  needed to isolate $\Omega_A$ from the rest of sites in the network. One may split $n(A)$ into two contributions, the $n(A)^{UV}\propto 2\log\, \beta/2\pi$ in the MERA curtain under consideration and a second one, $n(A)^{IR}$, accounting for the $\sim 2\pi(\ell/\beta)$ $i j$-bonds which must be cutted out in order to isolate $\sim 2\pi(\ell/\beta)$ $i$-sites from the rest of the network, each contributing $\log\, \chi$ to the entanglement entropy. Therefore,
 \begin{equation}
\label{eq2.12}
 S_A = \left[ n(A)^{UV} + n(A)^{IR}\right] \log\, \chi \sim  2\log\, \chi \, \left[ \frac{\pi \ell}{\beta} +  \log\,  \frac{\beta}{2\pi}\right]. 
\end{equation}

 One might demur on the apparently fine tuning of the bridge state $\vert \Psi \rangle$. In this sense, our guess for $\vert \Psi \rangle$ is done in order to match the behaviour of $S_A$ in a black hole geometry while establishing a reasonable simple relationship between the bond dimension $\chi$ of MERA and some conformal data of the boundary system.

\begin{figure}[t]
\label{bridge:mera}
\includegraphics[width=4.75 in]{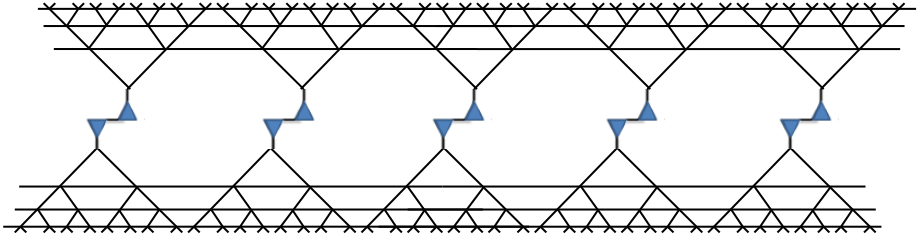}
 \centering \caption{A doubled MERA network with $\log \, \beta/2\pi$ layers of tensors associated at each MERA
 \emph{curtain} and an entangled bridge MPS  state $\vert \Psi \rangle$ gluing the two halves of the system.  Blue triangles with a downward leg represent $\lbrace i \rbrace$-sites while triangles with an upward leg correspond to $\lbrace j \rbrace$-sites. Thick lines between them represent entangled bonds of dimension $\chi$.}
\end{figure}

\section{Entanglement scrambling and black hole interiors}
\label{sec:scrambling}
Any purification of the reduced density matrix $\rho$ such as the one used to build the bridge MPS state $\vert \Psi\rangle=(\vert \psi \rangle_{ij})^{\otimes N}$, deals with an irreducible amount of arbitrariness in the following sense: from the point of view of the entanglement entropy between subsystems $\lbrace i \rbrace$ and $\lbrace j \rbrace$, this purification is completely equivalent to any state $\vert \Psi(\theta,\theta\, ')\rangle$ generated as,
\begin{equation}
\label{eq4.1}
\vert \Psi\rangle=(\vert \psi \rangle_{\lbrace i j \rbrace})^{\otimes N} \longrightarrow \mathcal{U}_{\lbrace i \rbrace}(\theta)\otimes  \mathcal{U}_{\lbrace j \rbrace}(\theta\, ')
(\vert \psi \rangle_{\lbrace i j \rbrace})^{\otimes N}=\vert \Psi(\theta,\theta\, ')\rangle,
\end{equation}
 where $\mathcal{U}_{\lbrace i \rbrace}(\theta)$ and  $\mathcal{U}_{\lbrace j \rbrace}(\theta\, ')$ are local unitary operations acting on $\lbrace i \rbrace$-sites and $\lbrace j \rbrace$-sites respectively and $\theta,\, \theta\, '$ parametrize these unitarities.
 
 As an example one might introduce forward time evolution on both sides ($\lbrace i \rbrace$ and $\lbrace j \rbrace$) of the bridge MPS state with $\theta=\theta\, ' = t$ , $\mathcal{U}_{\lbrace i \rbrace} = \exp(-i\, H_{\lbrace i \rbrace}^{u}\, t)$, $\mathcal{U}_{\lbrace j \rbrace} = \exp(-i\, H_{\lbrace j \rbrace}^{u}\, t)$ and $H = H_{\lbrace i \rbrace}^{u} = H_{\lbrace j \rbrace}^{u}$ being the coarse grained Hamiltonian of the original lattice system obtained under $u \sim \log \beta$ MERA renormalization steps. Although $\vert \Psi(\theta,\theta\, ')\rangle$ is not time translational-invariant, each individual density matrix $\rho$ on either side is time independent. Under this point of view, the thermofield double MERA state of the previous section might be considered as the tensor network configuration at $t=0$. As posed above, the tensor network at $t=0$ has a mass gap of the order of the temperature $T =2\pi/\beta$, which is a common feature of AdS black holes.  
 
 In the following, we show that the entanglement structures generated along the forward time evolution on both sides of the bridge state, have a tensor network representation \cite{Malda_TNS}. The effect of the unitary operations $\mathcal{U}_{\lbrace i \rbrace}(t)$ and  $\mathcal{U}_{\lbrace j \rbrace}(t)$ is to scramble the initially highly localized maximal entanglement structure between the $\lbrace i \rbrace$ and $\lbrace j \rbrace$ subsystems.  Namely, as $t$ grows, although the total amount of entanglement does not change, it spreads in such a way that, in general, it is not possible to describe the bridge state $\vert \Psi\rangle$ as a collection of unentangled $ij$-sites anymore. Despite this scrambling process, due to the entanglement invariance between the $\lbrace i \rbrace$ and $\lbrace j \rbrace$ subsystems under local unitary transformations, it is thus not possible to end up in any configuration in which the bridge state may be written as the product state $\vert \Psi \rangle = \vert \Phi\rangle_{\lbrace i \rbrace}\otimes \vert \Phi\rangle_{\lbrace j \rbrace}$. It will be shown below that, the full $\theta$-dependent unitary evolution can be interpreted in terms of a tensor network which grows along a spacelike $\theta$ direction, i.e, it is a tensor network which adds layers of $\vert \Psi(\theta,\theta') \rangle$ states forming a stack of bridge states.
 
\begin{figure}[t]
\includegraphics[width=2.5 in]{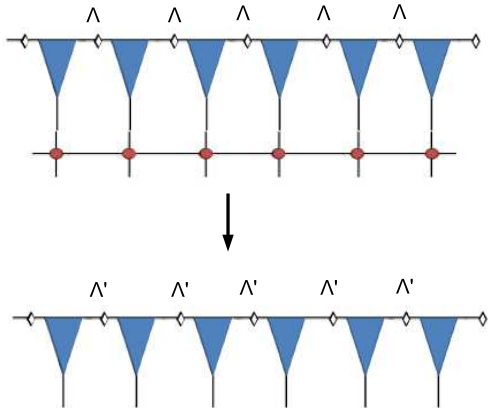}
 \centering \caption{A matrix product operator acting on the entire lattice. Red circles represent matrices $W$. The MPO is applied to an MPS by simply contracting the upper vertical legs with the physical indices of the state. After contracting the MPO with the initial MPS state, a new matrix product structure with increased bond dimension thus emerges.}
\end{figure}

\subsection{Local unitary evolution of the bridge state}
In Eq.(\ref{eq2.1}) it has been shown a matrix product representation of a reduced density operator in terms of matrices $M^{[k]}_{j_k\, j'_k}$ with two physical indices. Similarly, it has been shown in \cite{Cirac07} that, given a local Hamiltonian $H$, the infinitesimal time evolution operator $\mathcal{U}=\exp(-iH \delta t)$ can be given a similar matrix product operator (MPO) decomposition in terms of a set of matrices $W^{[k]}_{j_k\, j'_k}$ of dimension $d^2 \times d^2$. The beauty of this picture is that applying an MPO to an MPS, leaves the the MPS structure of the resulting state invariant, in other words, the resulting state can be written as an MPS in terms of a new set of matrices $\lbrace \Gamma',\, \Lambda'\rbrace$ at the cost of increasing the bond dimension $\chi$, i.e the dimension of the $\Lambda$ matrices, from $\chi \to d^2\, \chi$ (see Figure 5). Thus, the time evolution of any MPS state has a tensor network representation which dynamically encodes each resulting MPS state generated at each time step of the process.

 In the following, we elaborate on a tensor network description first proposed in \cite{Malda_TNS} which accounts for the scaling of the entanglement entropy of a quantum system during its thermalization after a quantum quench. To this aim, we present a procedure to independently time evolve both sides of the initial bridge state $\vert \Psi \rangle$  which generates two MPS states at each infinitesimal time step $\delta t$. The protocol has a tensor network representation depicted in Figure 6 and, assuming that both the MPO representations of the operators $\mathcal{U}_{\lbrace i \rbrace}(\delta t)$ and $\mathcal{U}_{\lbrace j \rbrace}(\delta t)$ are provided, it amounts to generate the doubled series of MPS states,
 \begin{eqnarray}
 \label{eq4.1b0}
 \vert \Psi(t + \delta t,0)\rangle = \mathcal{U}_{\lbrace i \rbrace}(\delta t)\, \otimes  \mathcal{U}_{\lbrace j \rbrace}(0)\, \vert \Psi(t,0)\rangle = \mathcal{U}_{\lbrace i \rbrace}(\delta t)\, \otimes  \mathbb{I}_{\lbrace j \rbrace}\, \vert \Psi(t,0)\rangle\\ \nonumber
 \vert \Psi(0,t + \delta t)\rangle = \mathcal{U}_{\lbrace i \rbrace}(0)\, \otimes  \mathcal{U}_{\lbrace j \rbrace}(\delta t)\, \vert \Psi(0,t)\rangle= \mathbb{I}_{\lbrace i \rbrace}\, \otimes  \mathcal{U}_{\lbrace j \rbrace}(\delta t)\, \vert \Psi(0,t)\rangle,
 \end{eqnarray}  
 each having the same time independent entanglement structure between $\lbrace i \rbrace$ and $\lbrace j \rbrace$ sites as the locally evolved states commented above. These states are bridge states in which the initially highly localized pattern of entanglement has been scrambled out. By carrying out a finite time evolution of this sort, its tensor network representation implictly builds up a full history of the process as it dynamically encodes each MPS state generated along it. It is clear that the structure of the network increases as time grows and eventually, the tensor network representation is equivalent to a stack of $\sim 2\, t$ MPS states.
 
 \begin{figure}[t]
\includegraphics[width=3.0 in]{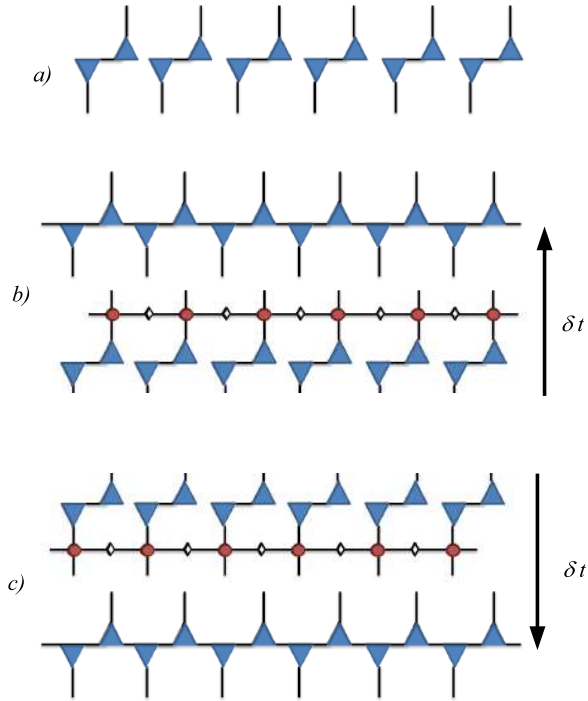}
 \centering \caption{a) The initial bridge MPS state. b) the MPO representation of $ \mathcal{U}_{\lbrace i \rbrace}(\delta t)$ acting on the initial bridge state generates a new bridge state $\vert \Psi(\delta t,0)\rangle$ in which the initially highly localized entanglement between $\lbrace i \rbrace$ and $\lbrace j \rbrace$ sites has been scrambled out so it cannot be written as $(\vert \psi \rangle_{ij})^{\otimes N}$ anymore. c) the MPO representation of $ \mathcal{U}_{\lbrace j \rbrace}(\delta t)$ acting on the initial bridge state yields another bridge state $\vert \Psi(0, \delta t)\rangle$.}
\end{figure}

\subsection{Linear growth of entropy from the black hole interiors}
\label{sec:bhinterior}
In this section, we analyze, following \cite{Malda_TNS}, the scaling of the entanglement entropy in a tensor network which  grows in a spacelike manner as $t$ grows. As commented above, this network adds layers of bridge states in the middle region within the two $u$-end layers of both MERA \emph{curtains}. In other words, the time evolution produces a wavefunction that may be represented in terms of a tensor network with a geometry which is simply longer. In some sense, the growing part of the tensor network acts as a record of the "history" of the state. As in \cite{Malda_TNS}, we interpret the middle region of the network, builded up as a stack of MPS, as the interior of the AdS black hole.  As we are implementing the forward $\theta$ evolution on both sides of the state commented above, we expect the number $N_{b}$ of these layers to be $N_{b} = 2\, \theta$ while we fix $\theta = 2\pi\, t/\beta$.

 We consider a region $A$ consisting of two identical disjoint pieces $A_1$, $A_2$ of $\ell \gg \beta$ sites, each one lying on the boundary of each of the two MERA \emph{curtains}. Then, we count the $n(A)$ bond indices which connect the region $\Omega_A$ with the rest of the sites in the network (Figure 7 left). While $t$ is small enough, this can be written as, 
 \begin{equation}
 \label{eq4.1b}
 n(A)^{t} = 4\log \frac{\beta}{2\pi} + 2N_{b} =  4\log \frac{\beta}{2\pi} + 8\pi \frac{t}{\beta}.
 \end{equation}

Since each bond contributes a maximum of $\log\, \chi$ to the entropy, the upper bound to $S_A$ is given by, 
 \begin{eqnarray}
 \label{eq4.2}
 S_A^{t}& = & 2S_{{\rm MERA}} + 8\pi \log \chi\, \frac{t}{\beta}, \\ \nonumber
 S_{{\rm MERA}}&=&2\log \chi\, \log \frac{\beta}{2\pi},
 \end{eqnarray} 
 which increases linearly in time as $t$ grows. For these regimes, $\Omega_A$ amounts to be a connected region inside the tensor network. Namely, $|\partial \Omega_A|$, the "extremal" minimal surface at early times, extends across the black hole interior from one asymptotic MERA \emph{curtain} to the other. This is consistent with the fact that, in holography, when considering static situations, the minimal surfaces do not penetrate into the event horizon, i.e, do not probe the black hole interior, but this interior can be probed as far as a time dependent system is under consideration. Thus, as time evolves, the bond counting progresses into the interior made of $N_b$ piled up MPS states, cutting two bonds for each state, each contributing, at most, $\log \chi$ to the entanglement entropy. Eventually, given Eq (\ref{eq1.1}), after some time evolution, the entropy of the region $A$ is better bounded by the thermal entropy of the two pieces of $\ell$ sites, which reads as,
\begin{equation}
 \label{eq4.2}
 S_A^{th} =  2S_{{\rm MERA}} + 4 \log \chi\, \frac{\pi\ell}{\beta}.
 \end{equation}
 
 Once this happens, the region $\Omega_A$ consists in two disconnected pieces as shown in Figure 7 (right). This transition in the structure of $\Omega_A$ has been also adressed in the context of AdS$_3$/CFT$_2$ and MERA tensor networks in \cite{headrick} and \cite{Molina11} respectively. Therefore, the entropy of region $A$ may be written as,
 \begin{equation}
 \label{eq4.3}
 S_A = {\rm min}\lbrace S_A^{th}, S_A^{t}\rbrace = \begin{cases}
 S_A^{t} = 2S_{{\rm MERA}} + \log \chi\, \left(8\pi\,  t /\beta\right), \, \quad t \leq \ell/2 \\
\\
S_A^{th} =  2S_{{\rm MERA}} + \log \chi\, \left( 4\pi\, \ell/\beta\right) \, \quad t > \ell/2
\end{cases}\,. 
 \end{equation} 
 
 Thus, after a time $t = \ell/2$, the extremal surface $|\partial \Omega_A|\sim n(A)$ splits off into two pieces. Both pieces stuck to each one of the two states $|\Psi(\ell/2, 0)\rangle$ and $|\Psi(0,\ell/2)\rangle$  yielded on both sides by forward time evolution, while $|\partial \Omega_A|$ becomes a static surface wich reproduces the thermal entropy. As a result, the entanglement grows linearly due to size increasing of the middle region along the $\theta$ direction, and eventually saturates at the thermal value $S_A^{th}$ at time $t \sim \ell/2$. The scaling of $S_A$ given in Eq. (\ref{eq4.3}) agrees with those obtained using direct CFT techniques \cite{calabrese} and holographic settings \cite{Malda_TNS}. One might interpret this fact in the following sense: after time $t \sim \ell/2$, sites in region $A$ are thermalized due to the spread of quantum correlations between its sites and the rest of the network.

 From equation (\ref{eq4.3}), it is easy to obtain the mutual information $\mathcal{I}(A_1:A_2)$ between the subsystems $A_1$ and $A_2$, which can be written as, 
\begin{eqnarray}
 \label{eq4.4}
 \mathcal{I}(A_1:A_2)& = & S(A_1) + S(A_2) - S(A_1\cup A_2)\\ \nonumber
&=& S_A^{th} - {\rm min}\lbrace S_A^{th}, S_A^{t}\rbrace = \begin{cases}
  \log \chi\, \left[8\pi /\beta \left( \ell/2 - t\right)  \right], \, \quad t \leq \ell/2 \\
\\
0 \, \quad t > \ell/2
\end{cases}\,, 
 \end{eqnarray}
 where  $S(A_1) = S(A_2)= 1/2\, S_A^{th}$ and $S(A_1 \cup A_2) = S_A$. The transition in the shape of $\Omega_A$ at time $t \sim \ell/2$ commented above is thus also responsible for the vanishing of the mutual information between the two regions. As the mutual information $\mathcal{I}(A_1:A_2)$ acts as an upper bound on the correlators between operators defined in those regions \cite{wolf},
\begin{equation}
\mathcal{I}(A_1:A_2) \geq \frac{(\langle {\cal O}_{A_1} {\cal O}_{A_2} \rangle - \langle {\cal O}_{A_1} \rangle \langle {\cal O}_{A_2} \rangle  )^2}{2|{\cal O}_{A_1}|^2 |{\cal O}_{A_2}|^2}~,
\label{eq4.5}
\end{equation}
 its vanishing signals an exponential decay of correlations between $A_1$ and $A_2$. Namely, in the AdS/CFT, this result arises because the RT formula only represents the leading term in a $1/N$ (or $G_N$) expansion.  The subleading terms in the long range expansion of the mutual information do lead to power-law corrections that in any case indicate that, for those regimes of separation, $\mathcal{I}(A_1:A_2)$ can be made parametrically small.   
 
 In \cite{swingle, vidal} it was pointed out that the behaviour of correlators in a tensor network (exponential in MPS \cite{mps_bib} and polynomial in MERA \cite{Vidal08}), is governed by the shape of geodesics in the discrete geometry of the tensor network. As commented above, once provided two sites $x_1$ and $x_2$ in a tensor network, then it is possible to define a distance between them, as far as they are connected through paths lying within the tensor network. Each one of these paths consists of a set of tensors and entangled bonds connecting the sites. One may associate a length to each path by simply counting the number of entangled bonds in the path. Finally, the distance $D(x_1,x_2)$ between $x_1$ and $x_2$ is defined by the length of the shortest path connecting them, i.e,
 \begin{equation}
 \label{eq4.5b}
 D(x_1,x_2) ={\rm min}\, \left\lbrace  n_{{\rm bonds}}( x_1, x_2)\right\rbrace .
 \end{equation} 
 
 It is worth to note that any "continuous" path within the network does not contain any unentangled bond. Indeed, an entangled bond between two adjacent sites is what precisely allow us to establish a notion of proximity between them. This  notion of distance relates with the possibility to adscribe a metric to the network \cite{ads_mera_tak} so, at least in the tensor network language, one might conclude that \emph{no entanglement means no geometry} \footnote{As has been pointed out in \cite{swingle, Malda_TNS}, each tensor may represent the wavefunction of a region of AdS radius, so this notion of proximity could be appropiate between regions which size is similar to the AdS radius but not for smaller distances.}.

Let us now consider the the correlation function of an operator $\mathcal{O}$ inserted at positions $x_1$ and $x_2$ in a tensor network, $\mathcal{C}(x_1,x_2)\equiv \langle \mathcal{O}(x_1)\, \mathcal{O}(x_2)\rangle$. The behaviour of $\mathcal{C}(x_1,x_2)$ for both the MPS and the scale invariant MERA may be written in terms of the distance $D(x_1,x_2)$ within the tensor network as,
\begin{equation}
	\mathcal{C}(x_1,x_2) \approx e^{-\alpha D(x_1,x_2)},
\label{eq4.6}
\end{equation}
with $\alpha$ some positive constant. As posed in \cite{vidal}, this is an asymptotic limit of the expression for the algorithmic computation of the correlator $\mathcal{C}(x_1,x_2)$ in the tensor network, which is given by,
\begin{equation}
	\mathcal{C}(x_1,x_2) \approx {\vec{v}_L}^{\, \dagger} \cdot  \mathbb{T}^{D(x_1,x_2)} \cdot \vec{v}_R. 
\label{eq4.7}
\end{equation}

This amounts to a scalar product of two vectors ${\vec{v}_L}$ and $\vec{v}_R$ with the $D(x_1,x_2)$-th power of a transfer matrix $\mathbb{T}$ from whose eigenvalues arise the correlation length $\xi$ for the gapped MPS or the power law for the scale invariant MERA (see \cite{mps_bib, Vidal08} for the original derivation of this result for MPS and MERA). 

 Let us now to consider the isertion at $t=0$ of two operators $\mathcal{O}$ at similar locations $x_1$ and $x_2$ within $A_1$ and $A_2$ respectively. After unitary time evolution of the middle region of the tensor network, it is thus expected that the minimal number of bonds connecting  $\mathcal{O}(x_1)$ and $\mathcal{O}(x_2)$ for $t\leq \ell/2$ is given by, 
 \begin{equation}
 \label{eq4.6}
 n_{{\rm bonds}}=2 \log\, \frac{\beta}{2\pi} + 4\pi\, \frac{t}{\beta}.
 \end{equation}
 
  Thus, assuming $n_{{\rm bonds}}$ amounts to a distance $D(x_1,x_2)$ within the discrete geometry of the tensor network, one obtains the exponential decay of $\mathcal{C}(x_1,x_2)$,
  \begin{equation}
 \label{eq4.7}
\mathcal{C}(x_1,x_2) \approx \exp \left(-\alpha\, n_{{\rm bonds}} \right) = \left(\frac{2\pi}{\beta} \right)^{2\alpha}\, e^{-4\pi\, \alpha\, t/\beta} ,
 \end{equation} 
 which characterizes the gapped nature of the middle region of the tensor network.
 
\begin{figure}[t]
\label{Linear:growth}
\includegraphics[width=4.0 in]{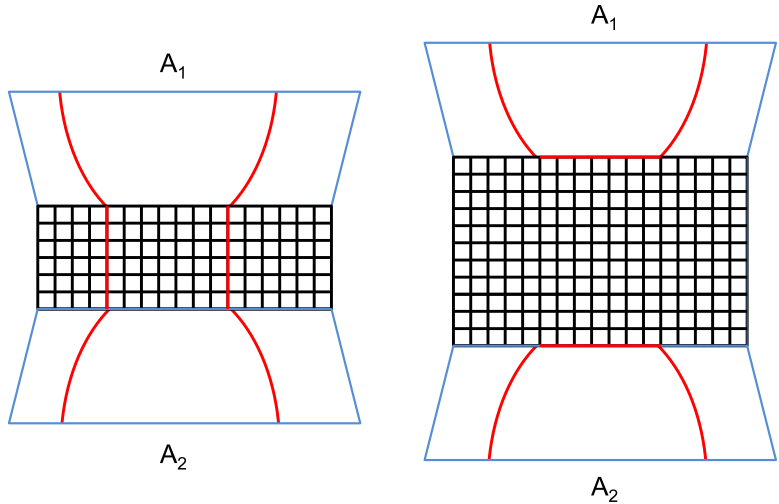}
 \centering \caption{The region $\Omega_A$ represents the reduced density matrix $\rho_A$ in a tensor network. Its boundary $\vert \partial \Omega_A \vert$ (red) scales as the number of entangled bonds $n_A$ that must be cutted to isolate the sites lying inside $\Omega_A$ from the rest of the sites of the network. While $t < \ell/2$, $\Omega_A$ is connected and its boundary is depicted in the leftmost figure. For $t \geq \ell/2$, there is a transition and $\Omega_A$ becomes disconnected as well as its boundary, as shown in the rightmost figure.}
\end{figure}
 
\subsection{Holographic interpretation}
\label{sec:holinterp} 
The structure of the middle region of the tensor network given by the forward time evolution on both sides of the bridge state, has been argued to be similar 
to the structure of the nice slices in the interior of the black hole \cite{Malda_TNS}. The gapped region represented by that tensor network is thus interpreted as the interior of the black hole in which each spacelike slice (a constant $t$ region from outside the horizon) corresponds to an MPS $\vert \Psi(t)\rangle$ state. In this sense, the full interior region of the black hole, including the horizon, is then represented by means of low energy eigenstates of a parent Hamiltonian. 

 This clearly differs from a time evolution of the MERA class of states which comprises both \emph{curtains} of the tensor network, which can be simulated, in principle, as proposed in \cite{montangero}. Now, the ansatz's tensors of MERA are the varying parameters along 
the evolution. In this case, it is clear that one evolves degrees of freedom outside the horizon. This approach has been addressed in terms of a continuous version of MERA in \cite{cMERA,Taka_MERA}. It would be interesting to compare both pictures in order to clarify whether the middle region of the tensor network detailed here, represents degress of freedom lying on the interior or just the region near, but outside the horizon.

 Nevertheless the interior interpretation of the middle region seems quite reasonable. In \cite{VanRaam} it has been argued that the structure of the entanglement between different parts of a system could act as a sort of glue from which a smooth classical spacetime between those regions emerges. Namely, in our tensor network construction, each MERA \emph{curtain} and its associated $\lbrace i\rbrace(\lbrace j \rbrace)$-sites correspond to the reduced density matrices of sets of complementary fundamental degrees of freedom lying at $A_1$, $A_2$. These density matrices might be interpreted as those from which one determines the value of observables within the left or right wedge of the eternal black hole (see Figure 8 left). The future interior region of the eternal black hole (upper part in Figure 8 left), i.e, the entanglement between $A_1$ and $A_2$ according to \cite{VanRaam}, is then encoded in the pattern of entanglement supported by the bridge MPS state. The non trivial forward time evolutions on both sides of the bridge state, grow the interior region, which amounts to modify the entanglement distribution (but not the amount) between the $\lbrace i\rbrace$ and $\lbrace j \rbrace$ sites of the state (see Figure 8 center). In this gravitational interpretation, the correlators in the right field theory are given by the right wedge while correlators in the left field theory are reproduced by the left wedge of the geometry. Similarly, in the tensor network, correlators on each side of the middle region are managed on each MERA \emph{curtain} separately. Furthermore, in the eternal black hole geometry, correlators with one leg on the left wedge and and the other one on the right wedge are specified through the future and past regions of the interior. In the tensor network description, these correlators are ruled out by the middle region, i.e, by the stack of bridge MPS states accounting for the entanglement distribution between the two MERA \emph{curtains}.

 Nevertheless, it is worth to note that the most general case, i.e, a global unitary operation of the bridge state,
 \begin{equation}
 \label{eq4.8} 
\vert \Psi\rangle=(\vert \psi \rangle_{ij})^{\otimes N} \longrightarrow \mathcal{U}_{\lbrace i\, j\rbrace}(\theta)\, (\vert \psi \rangle_{i j})^{\otimes N}=\vert \Psi(\theta)\rangle,
 \end{equation} 
  has not been adressed. Here, $\mathcal{U}_{\lbrace i\, j\rbrace}(\theta)$ amounts to a global unitary operation whose action, does not forbid to end up with a completely disentangled bridge state,
\begin{equation}
 \label{eq4.9} 
\vert \Psi(\theta)\rangle \sim \vert \Phi \rangle_{\lbrace i \rbrace}\otimes \vert \Phi \rangle_{\lbrace j \rbrace},
 \end{equation}
in which, all the initially localized entanglement has been removed. In this case, recalling the arguments of the previous section, one obtains a disconnected tensor network comprising two separated MERA \emph{curtains} from which neither a notion of a distance nor a geometry between the degrees of freedom of $A_1$ and $A_2$ can be established. This would correspond to excising the interior region of the black hole  (see Figure 8 right). 

\begin{figure}[t]
\label{vanraam:figure}
\includegraphics[width=4.75 in]{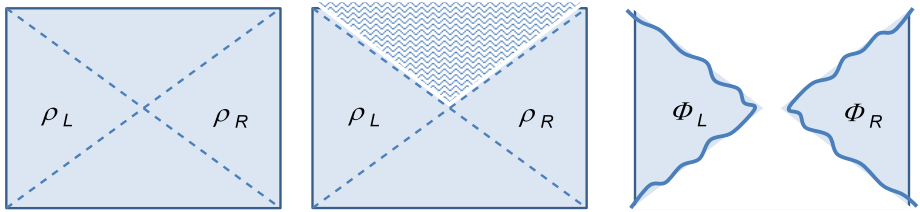}
 \centering \caption{Left: Connected AdS eternal black hole spacetime. Center: Connected AdS eternal black hole after local unitary evolution of the degrees of freedom lying inside the future interior region. Right: Two disconnected spacetimes from which a common interior region has been excised by carrying out a global unitary \emph{disentangling} operation.}
\end{figure}

\section{Conclusions} 
In this paper we have elaborated on the previous proposal in \cite{Malda_TNS} providing a tensor network in which the computation of entanglement entropy in certain situations remains consistent, at least qualitatively, with its holographic computation in the context of the eternal black hole in AdS/CFT. We have tried to emphasize that, in a tensor network, any two regions must be considered connected as far as there exist a continuous set of entangled bonds between them. This allows to establish a proto-notion of distance  within the network which could be useful in order to elucidate the proposed connection between entanglement (e.g, in tensor networks and quantum many body systems) and geometry \cite{Malda_Horizons}. It is also worth to investigate as well  if the knowledge of the structure of entanglement supported by systems with a suitable tensor network description, may clarify the role of large $N$ in a conjectured tensor network representation of a classical geometry. Presumably, this is related to the specific way in which the entanglement entropy bounds given by the network might be saturated in these systems. For this task, it could help to formulate the tensor network structures in terms of their continuous versions.

\acknowledgments
The authors thank S.R. Clark, J. Rodr\'iguez-Laguna and E. da Silva for giving very fruitful suggestions on the manuscript. JMV thanks the hospitality of Germ\'an Sierra and Esperanza L\'opez at the Institute of Theoretical Physics CSIC-UAM in Madrid. This work has been funded by Ministerio de Econom\'ia y Competitividad Project No. FIS2012-30625. 


\end{document}